\documentclass[twocolumn,pra,nofootinbib,showpacs,amsmath,amssymb,longbibliography]{revtex4-2}
\usepackage{times} 
\usepackage{braket}
\usepackage{babel}
\usepackage{blindtext}
\usepackage{physics}
\usepackage{color}
\usepackage{xspace}
\usepackage{amsmath}
\usepackage{amssymb}
\usepackage{amsbsy}
\usepackage{graphicx}
 \usepackage{bm}
 \usepackage{float}
\usepackage{relsize}
\usepackage[normalem]{ulem}
\usepackage{lipsum}

\usepackage[unicode,breaklinks]{hyperref}
\usepackage{scalerel}
\usepackage{tikz}
\usetikzlibrary{svg.path}

\definecolor{orcidlogocol}{HTML}{A6CE39}
\tikzset{
  orcidlogo/.pic={
    \fill[orcidlogocol] svg{M256,128c0,70.7-57.3,128-128,128C57.3,256,0,198.7,0,128C0,57.3,57.3,0,128,0C198.7,0,256,57.3,256,128z};
    \fill[white] svg{M86.3,186.2H70.9V79.1h15.4v48.4V186.2z}
                 svg{M108.9,79.1h41.6c39.6,0,57,28.3,57,53.6c0,27.5-21.5,53.6-56.8,53.6h-41.8V79.1z M124.3,172.4h24.5c34.9,0,42.9-26.5,42.9-39.7c0-21.5-13.7-39.7-43.7-39.7h-23.7V172.4z}
                 svg{M88.7,56.8c0,5.5-4.5,10.1-10.1,10.1c-5.6,0-10.1-4.6-10.1-10.1c0-5.6,4.5-10.1,10.1-10.1C84.2,46.7,88.7,51.3,88.7,56.8z};
  }
}

\newcommand\orcidicon[1]{\href{https://orcid.org/#1}{\mbox{\scalerel*{
\begin{tikzpicture}[yscale=-1,transform shape]
\pic{orcidlogo};
\end{tikzpicture}
}{|}}}}

\hypersetup{
    unicode=true,
    plainpages=false, 
    colorlinks=true,
    linkcolor=blue,
    citecolor=blue,
    filecolor=black,
    urlcolor=blue
}

\DeclareMathOperator{\erfc}{erfc}

\newcommand{\bw}{\mathbf{w}}
\newcommand{\bx}{\mathbf{x}}
\newcommand{\bk}{\mathbf{k}}
\newcommand{\Lsf}{\mathsf{L}}
\newcommand{\Xsf}{\mathsf{X}}
\newcommand{\ML}{\mathcal{L}}
\newcommand{\brho}{{\bm{\rho}}}
\newcommand{\bnu}{{\bm{\nu}}}

\begin{document}

\title{Excitations and phase ordering of the spin-stripe phase of a binary dipolar condensate}

\author{Au-Chen Lee \orcidicon{0000-0002-6384-090X}}
\author{D. Baillie \orcidicon{0000-0002-8194-7612}}
\author{P. B. Blakie \orcidicon{0000-0003-4772-6514}}

\affiliation{Department of Physics, Centre for Quantum Science,
and Dodd-Walls Centre for Photonic and Quantum Technologies, University of Otago, Dunedin, New Zealand}

\begin{abstract}
We consider the ground states, excitations  and dynamics of a quasi-two-dimensional binary dipolar Bose-Einstein condensate. Our focus is on  the transition to a spin-stripe ground state in which the translational invariance is spontaneously broken by a striped immiscible pattern of the alternating components. We develop a ground state phase diagram showing the parameter regime where the spin-stripe state occurs. Using Bogoliubov theory we calculate the excitation spectrum and structure factors. We identify a balanced regime where the system has a $\mathbb{Z}_2$ symmetry, and in the spin-stripe state this yields a nonsymmorphic symmetry. We consider the evolution of the system following a quench from the uniform to spin-stripe state, revealing novel ordering dynamics involving defects of the stripe order. Using an order parameter to characterize the orientational order of the stripes, we show that the phase ordering exhibits dynamic scaling.
\end{abstract}

\maketitle

\section{Introduction}

Recent experimental activity has revealed the tendency of suitably confined dipolar Bose-Einstein condensates (BECs) to form spatially structured states such a droplet crystals \cite{Kadau2016a} and supersolids \cite{Tanzi2019a,Bottcher2019a,Chomaz2019a}. These states occur in regimes where the system is unstable at the meanfield level, with beyond meanfield effects preventing mechanical collapse \cite{FerrierBarbut2016,Wachtler2016a,Bisset2016} (also see \cite{Petrov2015a}). Before these developments, theoretical proposals considered the possibility of structured ground states emerging in binary (i.e.~two-component) dipolar BECs \cite{Saito2009,Wilson2012a} (also see \cite{KuiTian2018}). Here the structure formation is driven by an instability to immiscibility, rather than collapse, and beyond meanfield effects are not required (cf.~\cite{Bisset2021,Smith2021a}). Interest in this system has increased with  the first experimental production of binary dipolar condensates of erbium and dysprosium atoms \cite{Trautmann2018} and the demonstration of interspecies Feshbach resonances  \cite{Durastante2020}.
The possibility of supersolidity in these systems has recently been theoretically explored in regimes where the components are miscible \cite{Scheiermann2023a,Halder2023b} and immiscible  \cite{Bland2022b,Li2022a,Halder2023b}, and a configuration supporting a self-bound (cohesive) crystals \cite{Arazo2023a} has been proposed.

Here we examine properties of immiscible crystal states occurring in a planar quasi-two-dimensional (quasi-2D) binary dipolar BEC. Our focus is on  stripe states with a one-dimension crystal pattern where a spatial modulation of the wavefunctions occurs along one direction in the plane of the trap. We refer to this as a spin-stripe state, because the broken translational invariance is most strongly revealed in the (pseudo)-spin density of the condensate, i.e.~the difference in density between the components. This work generalizes the recent immiscible supersolid states considered in cigar shaped potentials \cite{Bland2022b,Li2022a} to a quasi-2D planar case. In previous work \cite{Lee2022a} we have quantified the instabilities of this system in the uniform phase, identifying the conditions where density or spin modes cause dynamical instabilities and whether those modes are long-wavelength (phonon) or short-wavelength (roton) in character. In this paper we explore the regimes where uniform miscible state is unstable to a spin excitation resulting in immiscible ground states. 
For the spin-stripe ground states we calculate their properties and excitation spectrum. For excitations propagating along the planar direction normal to the stripes, the system has three gapless excitation branches and exhibits acoustic and optical phonon-like modes, as recently discussed by Kirkby \textit{et al.} \cite{Kirkby2023a}.  Here our focus is on the density- and spin-dynamical structure factors. We identify an interesting nonsymmorphic symmetry that is revealed by the excitations of the spin-stripe state when both components are suitably balanced.
 
 We also consider how the system orders into the spin-stripe phase following a sudden quench from the uniform miscible state. Immediately following the quench, we observe that small domains of spin-stripe order develop, choosing different orientations $\phi$ of the stripe pattern. The growth rate of the stripe-patterns is found to be in good agreement with the imaginary part of the frequency of the unstable spin-excitation mode of the initial state.  As time progresses we observe phase ordering dynamics, in which the spin-stripe domains  merge and grow.
 When the domains are relatively large compared to the microscopic length scales, we find that the domains grow with a power law, i.e.~$L_\phi(t)\sim t^{1/z}$, where $L_\phi$ is the domain size and $1/z$ is the dynamic critical exponent. In this regime we find that the order parameter correlation function displays dynamical scaling (i.e.~when lengths are scaled by $L_\phi(t)$, it is time-independent). While there has been some studies of the miscible to immiscible dynamics in finite binary dipolar BECs \cite{Wilson2012a,Smith2021a,Smith2023a,Halder2023a}, these have not considered a quantitative description of how the  order develops following such a quench. Furthermore, the presence of both superfluid and spin-stripe order makes this an interesting manybody system to study phase transition and ordering dynamics.

The outline of the paper is as follows.
In Sec.~\ref{Sec:Form} we develop the formalism for the ground-states and excitations of the planar 2D binary dipolar BEC. We also define the balanced regime, where the system has a $\mathbb{Z}_2$ symmetry between the components, which then manifests as a nonsymmorphic symmetry in the spin-stripe state. While our primary results are based on numerical solution of the associated Gross-Pitaevskii  and Bogoliubov-de Gennes (BdG) equations, we also develop a variational result for the spin-stripe state in the balanced regime.
In Sec.~\ref{Sec:Results} we present our results for the ground state phase diagram over a wide parameter regime. We find that spin-stripe phases are favored by a significant difference in the dipole strength of the components, so that a dipolar-nondipolar mixture or an antiparallel mixture (where the second component has opposite dipolar polarization to the first) favor the spin-stripe state. As the dipole-dipole interactions (DDIs) become more similar both within and between components, the spin-stripe lattice constant diverges and ground state approaches the usual case of an immiscible fluid  (i.e.~one large domain of each component). Our results for the density and spin dynamical structure factors illustrate the excitations of the spin-stripe state, and reveal the nonsymmorphic symmetry in the balanced case.
In Sec.~\ref{Sec:Qdynamics} we study the quench dynamics following a sudden change in parameters taking the system from a uniform miscible state into the regime where the ground state is a spin-stripe state. We examine the growth of local order   following the quench, and on longer time scales the spatial growth of order as signified by an order parameter characterizing the spin-stripe orientations. We show that at late times the order tends to grow with a power law and the dynamical scaling holds. We also observe defects of the order such as disclinations and grain boundaries. We conclude in Sec.~\ref{Sec:Conc}.

\section{Formalism}\label{Sec:Form} 
Our system is a two-component (binary) Bose gas of dipolar atoms with their magnetic moments polarized along the $z$-axis. These components could correspond to two different atoms or different spin states of a particular atom. In what follows we treat the masses of the atoms in each component having the same value $M$. The low energy interactions between these atoms are described by 
\begin{align}
U_{ij}(\mathbf{r})=g_{ij}^s\delta(\mathbf{r})+\frac{3g_{ij}^{dd}}{4\pi}\frac{1-3\cos^2\theta}{r^3},
\end{align}
where $i,j={1,2}$ label the components. Here $g_{ij}^s=4\pi a_{ij}^s\hbar^2/M$ is the $s$-wave coupling constant between components $i$ and $j$, with $a_{ij}^s$ being the $s$-wave scattering length. The  DDI coupling constant is $g_{ij}^{dd}=\mu_0\mu_i^m\mu_j^m/3$, with $\mu_i^m$ being the magnetic dipole moment along $z$ of component $i$. The DDIs are anisotropic with $\theta$ being the angle between the relative separation of the dipoles ($\mathbf{r}$) and the dipole polarization axis. Note that for the case of anti-parallel dipoles, e.g.~$\mu_1^m>0$ and $\mu_2^m<0$, we can have $g_{12}^{dd}<0$ while $g_{ii}^{dd}\ge0$ always.

We consider a system having strong axial harmonic confinement along the $z$-axis with frequency $\omega_z$, and unconfined in the $xy$-plane, i.e. $V(z)=\frac{1}{2}M\omega_z^2z^2$.

\subsection{Meanfield ground states}\label{Sec:Grndstates}
The meanfield description for the ground state wavefunctions $\Psi_i(\bx)$ of a zero temperature BEC is provided by the two-component dipolar Gross-Pitaevskii  theory \cite{Goral2002} with energy functional
\begin{align}
E&=\int d\bx\,\sum_{i=1}^2 \Psi_i^*(\bx)\bigg[-\frac{\hbar^2\nabla^2}{2M}+V(z)\bigg]\Psi_i(\bx) \label{EGPE3D}\\
&+\frac{1}{2}\sum_{i,j=1}^2\int d\bx\,d\bx^\prime\,|\Psi_i(\bx)|^2U_{ij}(\bx-\bx^\prime)|\Psi_j(\bx^\prime)|^2 . \nonumber
\end{align}
We assume the system is in the quasi-2D regime, where interaction energy scales are small compared to $\hbar\omega_z$ so that axial degrees of freedom are frozen out and the axial wavefunction is well-approximated by the harmonic oscillator ground state 
$\chi(z)=e^{-z^2/2l_z^2}/\big(\pi^{1/4}\sqrt{l_z}\big)$,  where $l_z=\sqrt{\hbar/M\omega_z}$ is the harmonic oscillator length. 
We search for stripe phases, with variation along only one direction in the plane, which we take to be the $x$ direction, giving a ground state wavefunction of the form
\begin{align}
 \Psi_i(\bx)=\sqrt{n}\psi_i(x)\chi(z), \label{wfn3D} 
\end{align}
where we have taken the areal density, $n$, of each component to be the same, and $\psi_i(x)$ is dimensionless. This ansatz allows us to consider uniform miscible states with $\psi_i(x)=1$, and striped states for which we take the spatial variation to be periodic with unit cell (uc) length $L$, i.e.~$\psi_i(x)=\psi_i(x+L)$, with normalization constraint $\int_{\text{uc}} dx|\psi_i(x)|^2=L$. In practice a slight tilt of the dipole polarization to have a transverse component could be used to give preference to a particular direction (e.g.~see \cite{Wilson2012a}). 
We focus our attention here on striped states since two-dimensional (2D) patterns are generally of higher energy for the parameter regimes we consider.

Our interest is to determine the wavefunction that minimizes the energy per particle for fixed average areal density $n$. We can do this by finding the energy Eq.~\eqref{EGPE3D} in a single unit cell and dividing by the number of particles in the cell, $N$ to give 
\begin{align}
    \mathcal{E}_L 
    &=\sum_{i=1}^2\int_\text{uc} \frac{dx}{2L}\,\psi_i^*\Bigg[-\frac{\hbar^2}{2M}\frac{d^2}{dx^2}+ \frac{1}{2}\sum_{j=1}^2\Phi_{ij}(x)\Bigg]\psi_i,\label{EGPE2D}
\end{align}  
which is the energy per particle.
Here, we have neglected the constant zero-point energy of $\chi$, and the interactions are described by 
\cite{Ticknor2011a,Wilson2012a,Baillie2015a,Lee2022a}
\begin{align}
\Phi_{ij}(x)={n}\mathcal{F}^{-1}\!\Big\{\tilde{U}_{ij}(k_x)\mathcal{F}\!\left\{|\psi_j|^2 \right\}\!\Big\},
\end{align} 
where $\mathcal{F}$ denotes the one-dimensional Fourier transform of the $x$ coordinate in a unit cell and $\mathcal{F}^{-1}$ is the inverse transform (against the $k_x$ coordinate). Here the quasi-2D interaction kernel is
 \begin{align} 
 \tilde{U}_{ij}(k_x)&=\frac{g^s_{ij}+g_{ij}^{dd}\big[2-3G_0\big(k_x l_z/\sqrt{2}\big)\big]}{{\sqrt{2\pi}l_z}},
 \end{align}
 with 
$G_0(q)=\sqrt{\pi}qe^{q^2}\!\erfc(q)$, where $\erfc$ is the complementary error function.
 
 To determine the ground states we minimize  $\mathcal{E}_L$ with respect to the wavefunctions $\{\psi_1,\psi_2\}$ for any given $L$. Then find which uc length $L$ gives the minimum $\mathcal{E}_L$. For the uniform miscible case the wavefunctions are constant and  $\mathcal{E}_L$ is independent of $L$.  In general the stationary state solutions 
  will satisfy the coupled Gross-Pitaevskii equations
 \begin{align}
\mathcal{L}_{\mathrm{GP},i}\psi_i(x) = \mu_i\psi_i(x) ,\label{tiGPE}
 \end{align}
 where $ \mathcal{L}_{\mathrm{GP},i}=-\frac{\hbar^2}{2M}\frac{d^2}{dx^2}+  \sum_{j}\Phi_{ij}(x)$ and $\mu_i$ is the chemical potential for component $i$.
In practice we locate these solutions using gradient flow (imaginary time) evolution (e.g.~see \cite{Blakie2020b,Lee2021a,Smith2023a}) to optimize the wavefunctions.
 
\subsection{Excitations and structure factors}
The collective excitations can be obtained by linearizing the time-dependent GPE, $i\hbar\dot{\Psi}_j = \delta E/\delta \Psi^*_j$ around a stationary solution as
\begin{equation}
\Psi_j(\bx,t)=e^{-i\mu_j t/\hbar}\left[\sqrt{n}\psi_j(x)+\vartheta_j(\brho,t)\right]\chi(z),
\end{equation}
where $\brho =(x,y)$. From the ground state symmetries the excitations take the form
\begin{align}
\vartheta_j(\brho,t)\equiv\sum_\bnu\bigl[
 &c_\bnu   u_{\bnu j}   e^{i(q_xx+k_yy-\omega_\bnu t)} \notag\\
-&c_\bnu^* v_{\bnu j}^*  e^{-i(q_xx+k_yy-\omega_\bnu^* t)}
\bigr],\label{Eq:excitation}
\end{align} 
where $c_{\bm{\nu}}$ are the (assumed small) expansion coefficients, $\{u_{\bnu j}(x) ,v_{\bnu j}(x) \}$ and $\omega_{\bm{\nu}} $ are the quasi-particle amplitudes and angular frequencies respectively. The quantum numbers to describe the excitation, $\bnu =(\nu,q_x,k_y)$, with $|q_x|\le K/2$ being the quasi-momentum in the first Brillouin zone, where $K=2\pi/L$ is the reciprocal lattice vector, $k_y$ describing the planewave behavior along $y$, and 
$\nu$ representing the remaining band information.
We have used a Bloch form for the $x$-dependence of the wavefunction, with $\{u_{\bnu j}(x),v_{\bnu j}(x)\}$ being periodic functions on the unit cell. 
 
The excitations satisfy the BdG equations  $H_{q_x,k_y}\bw_\bnu = \hbar\omega_\bnu\bw_\bnu$ where $\bw_\bnu = (u_{\bnu1},u_{\bnu2},v_{\bnu1},v_{\bnu2})^T$ and 
\begin{align}
H_{q_x,k_y}&=\begin{pmatrix}
\Lsf+\Xsf&-\Xsf\\ 
\Xsf&-\Lsf-\Xsf
\end{pmatrix}.\label{BdG}
\end{align} 
Here we have introduced the matrices $\Lsf$ with components $L_{ij}= (-1)^{i-1}\ML_i\delta_{ij}$ where 
\begin{align}
\mathcal{L}_j &=\frac{\hbar^2}{2M}\left[\left(q_x-i\frac{d}{dx}\right)^2+k_y^2\right]+\sum_{k=1}^{2}\Phi_{jk}(x)-\mu_j, \label{Eq:BdG_L}
\end{align}
and $\mathsf{X}$ with components
\begin{align}
  X_{ij}f &= n\psi_i\mathcal{F}^{-1}\!\bigg\{\tilde{U}_{ij}\bigg[\sqrt{(q_x+k_x)^2+k_y^2}\bigg]\mathcal{F}\!\left\{\psi_jf \right\}\!\bigg\}.
\end{align}
We can quantify the  character of the excitations by introducing dynamic structure factors, which can be measured through Bragg spectroscopy \cite{Bismut2012a,Petter2019,Petter2021a,Baillie2016a}.  In a two-component system we can define a \textit{density}  ($+$) and the \textit{spin-density} ($-$) dynamic structure factor as \cite{Lee2021b} 
\begin{align}
    S_\pm(\bk_\rho,\omega)=\sum_{\nu}|\delta\tilde{n}_{\bk_\rho\nu}^\pm|^2\delta(\omega-\omega_{\bm{\nu}}),\label{Eq.SF}
\end{align}
where $\bk_\rho=(k_x,k_y)$. The wavevector $k_x$ determines the quasimomentum $q_x$, reduced to the first Brillouin zone by the reciprocal lattice vector $K$ for integer $m$, i.e. $k_x=q_x+mK$. The density fluctuations are $\delta n_{\bnu j}(x)e^{i(q_xx+k_yy)}$ where 
\begin{align}
\delta n_{\bnu j}(x)&=\sqrt{n}\psi_j(x)[u_{\bnu j}(x)-v_{\bnu j}(x)],\\
    \delta\tilde{n}_{\bk_\rho\nu}^\pm 
    &=\int_\mathrm{uc} d\brho \,e^{-imKx}[\delta n_{\bnu1}(x)\pm\delta n_{\bnu2}(x)].\label{Eq:DenFlu_FFT}
\end{align}
Consistent with our quasi-2D approximation, we neglect axial excitations\footnote{This is a reasonable approximation as only even bands can contribute to the $ S_\pm$, and their contribution will generally be weak.}, which will appear for  $\omega\gtrsim\omega_z$.

\subsection{Balanced regime}
We now specialize the formalism  outlined in the previous subsections to the balanced case -- where, in addition to the areal density of the components being identical, we take the magnitude of the dipole moments, and the intraspecies contact interactions of each component to be the same, i.e.
\begin{align}
\begin{array}{ll}
g_{11}^s&=g_{22}^s,   \\[2\jot]
g^{dd}_{11}&=g^{dd}_{22}=|g^{dd}_{12}|.
\end{array}\quad\text{(balanced regime)}
\end{align} 
The latter condition requires the magnitude of the magnetic moments of both species to be identical, but allows for the moments to be aligned parallel ($\mu_1^m=\mu_2^m$, $g^{dd}_{12}>0$) or antiparallel ($\mu_1^m=-\mu_2^m$, $g^{dd}_{12} = -g^{dd}_{ii}<0$). Under these conditions the Hamiltonian has a $\mathbb{Z}_2$ symmetry, i.e.~invariance to exchange of the components.

\subsubsection{Ground symmetry and variational approach}
Motivated by our numerical results in this regime [e.g.~see  Fig.~\ref{Fig:USPtransition}(c1)], the ground state wavefunctions of both components can be chosen to be real, with the wavefunction of one component equal to the other translated by half a unit cell, i.e.~
\begin{align}
	\psi_1(x)=\hat{T}_{L/2}\psi_2(x),\label{balancedwfns}
\end{align}
where $\hat{T}_{a}$ is the operator implementing a spatial translation of $a$ along $x$.  This symmetry manifests in the results for the excitations we present later.

Due to the similarity of both components we describe the spin-stripe state, with a variational wavefunction of the form 
\begin{align}
\psi_i(x)= \cos\eta+(-1)^{i-1}\sqrt{2}\sin\eta\cos(2\pi x/L) .\label{varwfn}
\end{align}
Here $\eta$ is the order parameter for the spin-stripe, with the state being uniform for $\eta=0$.
This wavefunction satisfies Eq.~(\ref{balancedwfns}) and the normalization condition $\int_\mathrm{uc} dx|\psi_i|^2=L$.
We restrict $\eta\in[0,\eta_0]$, where $\eta_0=\cot^{-1}\sqrt{2}$ is the value at which the wavefunction has a zero crossing.  Evaluating the energy per particle \eqref{EGPE2D} we obtain
\begin{align}
&\mathcal{E}_\mathrm{var}(L,\eta)=\frac{h^2\sin^2\eta}{4ML^2} \label{varEfun}\\
&+\frac{n}{4}\left[\tilde{U}_{+}(0)+  \sin^22\eta\tilde{U}_{-}\left(\frac{2\pi}L\right)+\frac{1}{2}\sin^4\eta\tilde{U}_{+}\left(\frac{4\pi}L\right)\right], \nonumber
\end{align}
where  $\tilde{U}_\pm(k)=\tilde{U}_{ii}(k)\pm\tilde{U}_{12}(k)$. Variational solutions can be identified by finding the minima of $\mathcal{E}_\mathrm{var}(L,\eta)$.

\subsubsection{Excitations}\label{Sec:BalanceExcitations}
We can formalize the symmetry observed in Eq.~(\ref{balancedwfns}) with the operator  
\begin{align}
\Lambda= T_{L/2}  \sigma_x,
\end{align}
  where $\sigma_x= \begin{pmatrix} 0 & 1 \\ 1 & 0\end{pmatrix}$ is the Pauli $x$ matrix acting in the pseudo-spin-$\tfrac12$ space of the components. 
The GPE and BdG operators respect the $\Lambda$ symmetry, i.e.~
 \begin{align} 
H_{q_x,k_y} &= 
 \begin{pmatrix}
\Lambda & 0\\
0 & \Lambda
\end{pmatrix}H_{q_x,k_y} 
 \begin{pmatrix}
\Lambda & 0\\
0 & \Lambda
\end{pmatrix}^\dagger,
 \end{align}\\
 and the excitations can be taken to be eigenstates of $\Lambda$ as
 \begin{align}
 \begin{pmatrix}
 	\Lambda & 0\\
 	0 & \Lambda
 \end{pmatrix} \bw_\bnu  =\lambda \bw_\bnu.\label{Eq:uv_nonsymm}
 \end{align}
Here we also introduce the quantum number $\lambda=\pm1$. In the uniform state the symmetry is trivially reduced to $\sigma_x$, and reflects that excitations can be chosen to be in-phase ($\lambda=1$, \textit{density}) modes or out-of-phase ($\lambda=-1$, \textit{spin}) modes (see Ref.~\cite{Lee2022a}). For the modulated case $\Lambda$ reflects a nonsymmorphic symmetry,  which arises from a combination of point-group operations with nonprimitive lattice translations (see
Ref.~\cite{Zhao2016a}).

\section{Results for ground states and excitations}\label{Sec:Results}

\subsection{Ground state phase diagram}
 \subsubsection{Uniform miscible state instabilities} 
  
    \begin{figure}[htp!] 
	\centering
	\includegraphics[trim=0 0 0 0,clip=true,width=1.02\linewidth]{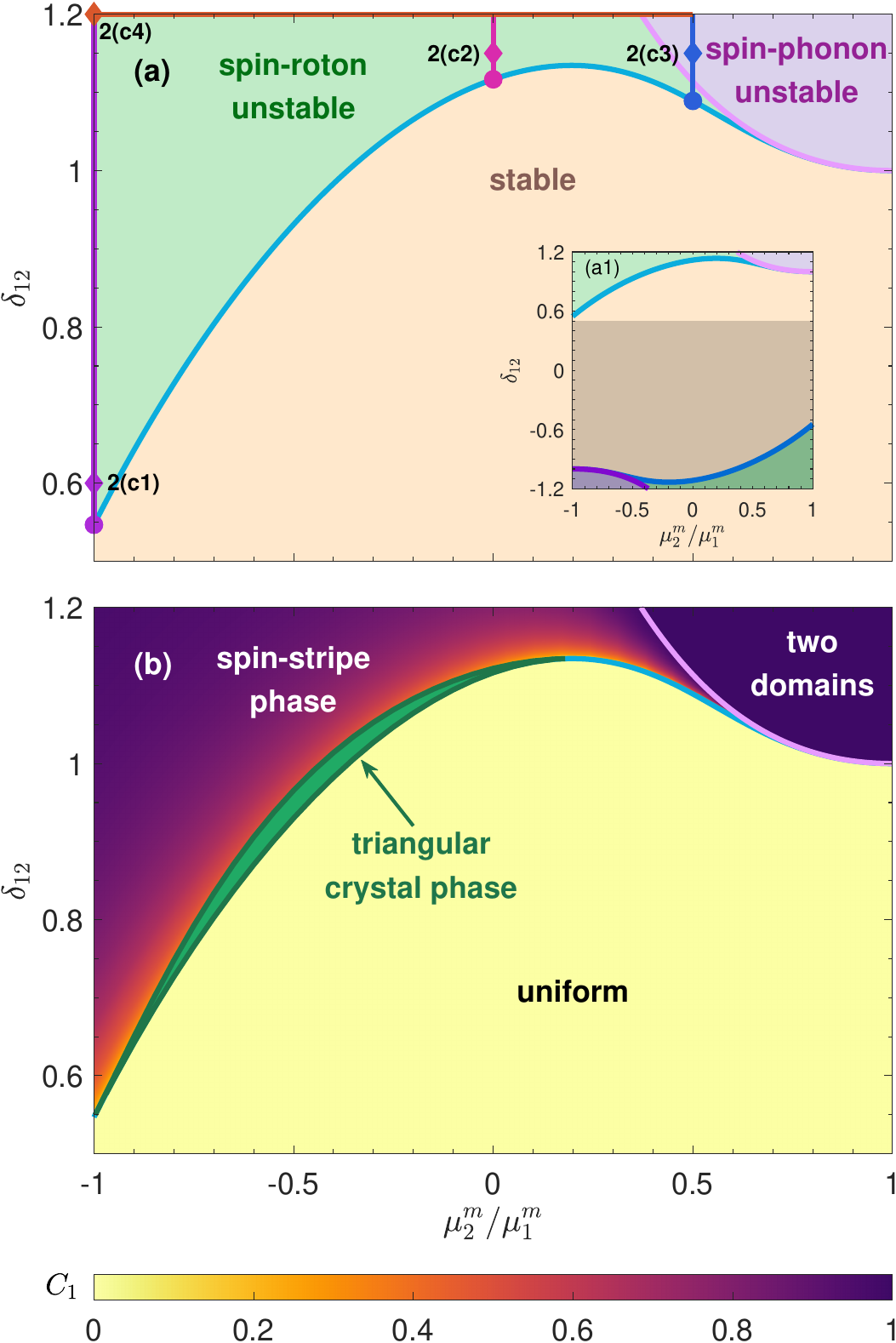}
	\caption{(a) Dynamical stability diagram  of a miscible uniform state. Inset (a1) shows the stability diagram over a broader parameter regime (details of instability identification can be found in \cite{Lee2022a}). Colored lines and attached markers in (a) correspond to the results shown in Fig.~\ref{Fig:USPtransition} with the same color. (b) Ground state phase diagram. Contrast $C_1$ is used as the spin-stripe order parameter, except in the green shaded region, where the ground state has a triangular crystalline pattern.    In the two domains region we have set $C_1=1$. Here $C_1\to1$ but it is difficult to calculate accurately due to the large domain size and sharp boundaries [also see Figs.~\ref{Fig:USPtransition} (b2) and (c3)].  Calculation parameters:   $ng^s_{ii}/\hbar\omega_zl_z=5$ and $g_{11}^{dd}/g^s_{11}=0.9$.
	}\label{Fig:PDiag}
\end{figure}
 
 First we consider the system in a uniform miscible state  (i.e.~$\psi_i=1$) and study its collective excitations to quantify when it becomes unstable. 
 Our results for this are shown in Fig.~\ref{Fig:PDiag}(a) as the relative dipole moment and the inter-species contact interaction vary. Here we have described the interspecies interaction in terms of the parameter 
 \begin{align}
 \delta_{12}=\frac {g_{12}^s}{\sqrt{g_{11}^sg_{22}^s}}.
 \end{align}
For reference, a non-dipolar binary condensate is stable and miscible for $|\delta_{12}|\le1$ and is immiscible for $\delta_{12}>1$.  
  
The dynamical stability diagram presented in Fig.~\ref{Fig:PDiag}(a) is similar to that explored in Ref.~\cite{Lee2022a}, but here is specialized to the quasi-2D assumption and is restricted to where immiscibility transitions occur\footnote{We exclude the region of mechanical instability which requires accounting for quantum fluctuations (e.g.~see \cite{Bisset2021,Smith2021a}).}. The two unstable regions in Fig.~\ref{Fig:PDiag}(a) correspond to where long wavelength spin phonon modes or  short wavelength spin roton modes soften and become dynamically unstable.  The spin character of the unstable modes can be assessed from their dominant contribution occurring in the $S_-$  structure factor (see Fig.~\ref{excitationsfig0} and \cite{Lee2022a}) and for the balanced case these modes have $\lambda=-1$.
Such unstable modes suggest that the components will spatially separate leading to an immiscible transition. This indicates an immiscible state will emerge as the ground state.

   \begin{figure}[htp!] 
	\centering
	\includegraphics[trim=0 0 0 0,clip=true,width=1.02\linewidth]{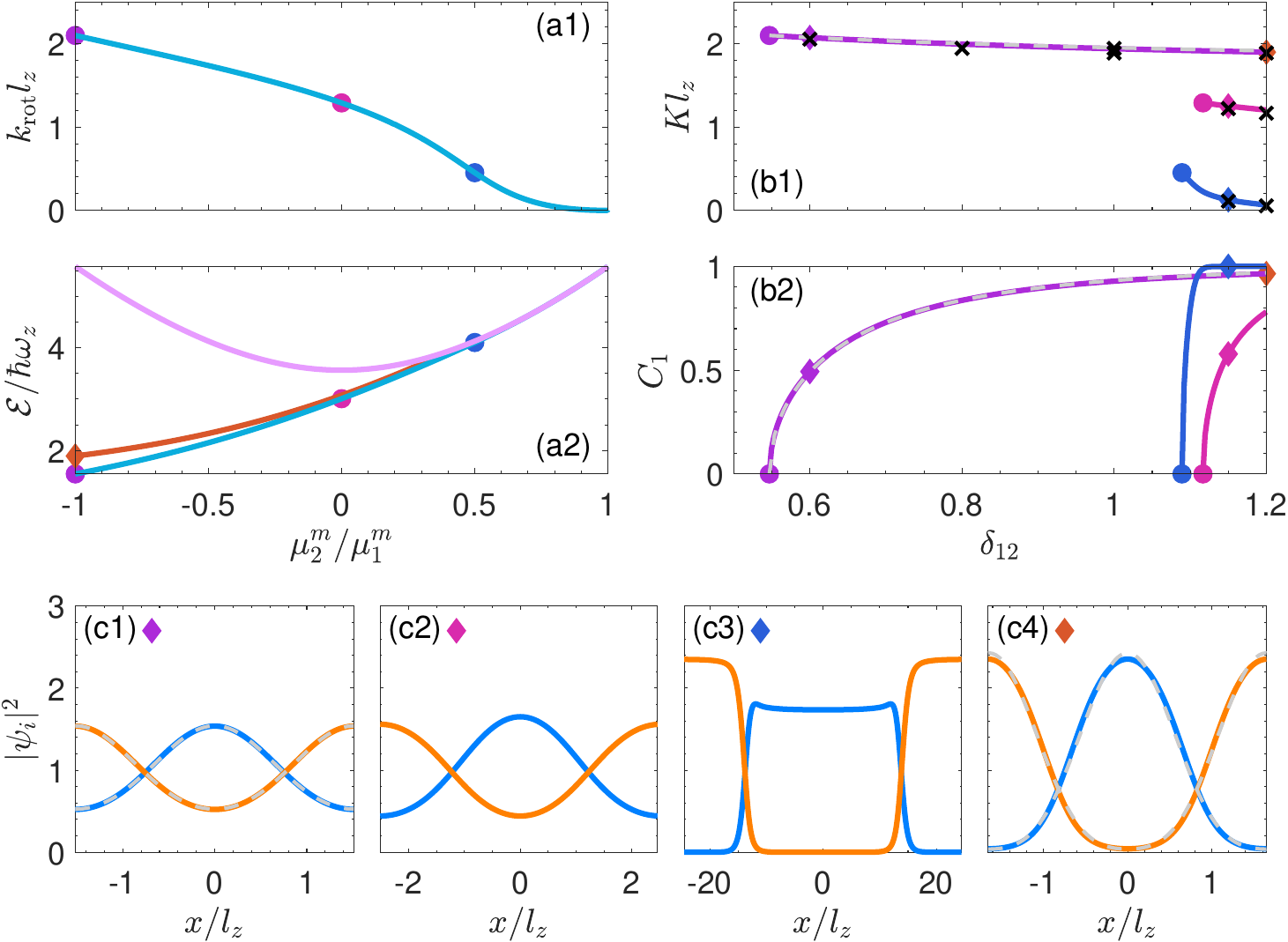}
	\caption{System properties along specified lines in the dynamic stability diagram [Fig.~\ref{Fig:PDiag}(a)]: (a1) Roton wave vectors. (a2) Energy per atom. (b1) Reciprocal lattice vectors. (b2) Contrast of component 1. 
	(c1)-(c4) Spin-stripe state densities of component $1$ (blue) and $2$ (orange).  Colored lines and markers in (a) and (b) are for the parameters indicated by the same lines and markers in Fig.~\ref{Fig:PDiag}(a).  The diamonds labeling the (c) subplots identify the solution parameters in Fig.~\ref{Fig:PDiag}(a). Gray dashed lines in (b1), (b2), (c1), and (c4) are variational results for balanced case. Black cross markers in (b1) are the average domain size from dynamical results from the simulations presented in Fig.~\ref{earlytime}(b).
	}\label{Fig:USPtransition}
\end{figure}

\subsubsection{Ground state phase diagram}
 Using the formalism outlined in Sec.~\ref{Sec:Grndstates} we can find the ground states in the regions where the uniform miscible state is dynamically unstable. The ground state phase diagram is shown in Fig.~\ref{Fig:PDiag}(b),  with some example  states shown in   Figs.~\ref{Fig:USPtransition}(c1)-(c4).  
When the uniform miscible state is unstable, the ground state has a modulated density in-plane, arising because the two components are immiscible and partially separated.  The modulation of the density of component 1 is characterized by the density contrast 
 \begin{align}
 	C_1=\frac{\max|\psi_1|^2-\min|\psi_1|^2}{\max|\psi_1|^2+\min|\psi_1|^2}.
\end{align}  
Here a contrast of 0 indicates that the state is uniform, and a (maximal) contrast of 1 indicates that the density of that component goes to zero at some point in the unit cell. 
We shade the phase diagram in Fig.~\ref{Fig:PDiag}(b) according to the component-1 density contrast of the lowest energy spin-stripe state. 
Note that when component 1 begins to modulate (i.e.~$C_1>0$) then  component 2 also modulates such that the dense regions of each component avoid each other [e.g.~see Figs.~\ref{Fig:USPtransition}(c1)-(c4)].    

For parameters considered in our results, the transition from the uniform to spin-stripe state is continuous, i.e.~the contrast emerges continuously as $\delta_{12}$ varies  [see Fig.~\ref{Fig:USPtransition}(b2)]. 
The reciprocal lattice vector of the spin-stripe state decreases with $\delta_{12}$ [see Fig.~\ref{Fig:USPtransition}(b1)], but most importantly it is strongly dependent on the relative dipole strength between the components, as characterized by the ratio $\mu^m_2/\mu^m_1$ [see Fig.~\ref{Fig:USPtransition}(a1)].
  In the region where the uniform state exhibits a spin-roton instability [light green shaded region in Fig.~\ref{Fig:PDiag}(a)], the stripe state emerges with microscopic length scale  $L\sim l_z$. For example, $L\approx\pi l_z$ for the anti-parallel case, increasing to  $L\approx1.6\pi l_z$ for the dipolar-nondipolar case.
   In the region where the uniform state exhibits a long wavelength spin-phonon instability   [lavender shaded region in Fig.~\ref{Fig:PDiag}(a) and as $\mu_1^m/\mu_2^m\to1$ in Fig.~\ref{Fig:USPtransition}(a1)] the length $L$ diverges, consistent with the system preferring to become immiscible by forming a single large domain of each component. We refer to this as the two domains region, which can be understood using a model with two uniform condensates of equal average areal density $n$ and no inter-component interaction. In the ground state of this model, the two components have equal pressure and the energy per particle \eqref{EGPE2D} is
\begin{align}
	\mathcal{E}_{\mathrm{LD}}=\frac{n}{4}\bigg(\sum_{i=1}^2\sqrt{g_{ii}^s+2g_{ii}^{dd}}\bigg)^2,
\end{align}
shown as a pink line in Fig.~\ref{Fig:USPtransition}(a2). Equating this to the energy of a miscible uniform case in the quasi-$2$D approximation [Eq.~\eqref{varEfun} with $\eta=0$] yields an result coinciding with the BdG approach used  in \cite{Lee2022a} and shown as the pink line in Figs.~\ref{Fig:PDiag}(a) and (b).

We also indicate in Fig.~\ref{Fig:PDiag}(b) the small region where a triangular immiscible state emerges as the ground state via a first order transition from the uniform state. The roton softening boundary [light blue solid line in Figs.~\ref{Fig:PDiag}(a) and (b)] is enclosed within it, i.e.~the transition to the triangular state generally occurs at a slightly lower $\delta_{12}$ value. Stronger nonlinearities  (i.e.~larger $ng_{ij}^s$ and $ng_{ij}^{dd}$), more strongly imbalanced regimes, or in-plane confinement could be used to favor the triangular spin state over a broader parameter regime, but we do not consider this further here.

For the balanced anti-parallel case $\mu_1^m/\mu_2^m=-1$ we compare the numerical and variational results for the spin-stripe state properties in Figs.~\ref{Fig:USPtransition}(b1), (b2), (c1), and (c4). These comparisons show an excellent agreement and verify the utility of the variational approach to the ground states in the balanced regime.

\begin{figure}[htp!] 
	\centering
	\includegraphics[trim=0 0 0 0,clip=true,width=\linewidth]{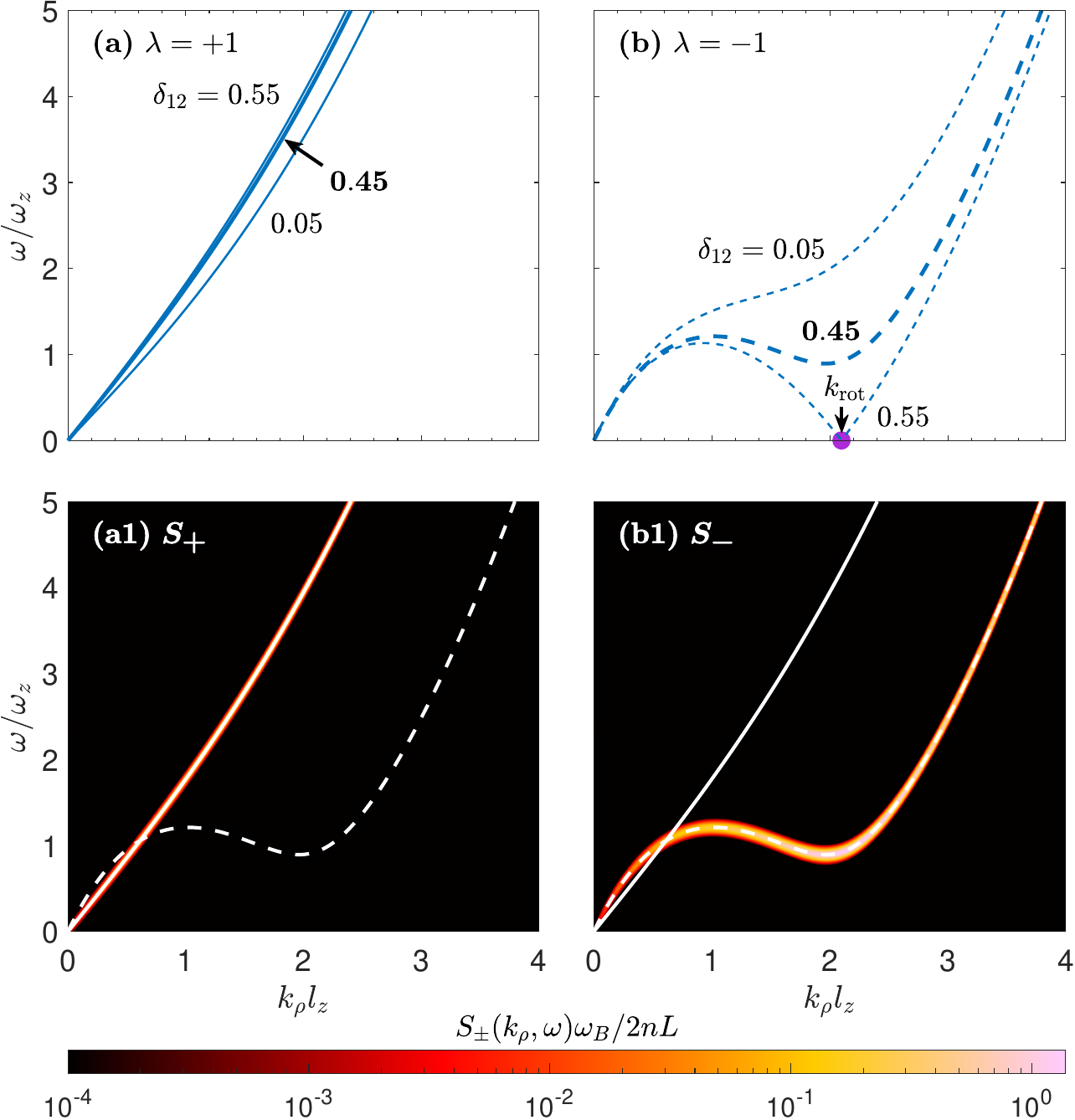}
	\caption{
	Excitations and dynamic structure factors for balanced anti-parallel uniform states close to the spin-stripe transition.
    Excitations with (a) $\lambda=1$  and (b) $\lambda=-1$  for three values of $\delta_{12}$. (a1) Density and (b1) spin dynamic structure factors for the uniform ground state with $\delta_{12}=0.45$. The relevant $\lambda=1$ (solid) and $\lambda=-1$ (dashed) excitations branches are shown in (a1) and (b1) as white lines for reference. Excitations for $\delta_{12}=\delta_c=0.55$ shown in (a) and (b) correspond to the purple circles in Figs.~\ref{Fig:PDiag}(a) and \ref{Fig:USPtransition}(a,b). The dynamical structure factors were broadened by setting $\delta(\omega)\to e^{-(\omega/\omega_B)^2}/\sqrt{\pi}\omega_B$ in Eq.~(\ref{Eq.SF}) with $\omega_B=0.04\omega_z.$
	\label{excitationsfig0} }
\end{figure}

\subsection{Excitations and structure factors}
 \subsubsection{Uniform miscible state}
In  Figs.~\ref{excitationsfig0}(a) and (b), we consider the excitations in uniform balanced ground states with anti-parallel dipoles $\mu^m_1/\mu_2^m=-1$ for several values of $\delta_{12}$. In the uniform state  there are two gapless excitation branches. For the balanced case, these branches have density ($\lambda=1$) and spin ($\lambda=-1$) character.
As $\delta_{12}$ increases, the slope of the density branch near $k_\rho=0$  (where $k_\rho^2= k_x^2+k_y^2$) also increases.  In contrast the spin excitation branch develops a roton-like (local-minimum) feature that softens to zero energy as $\delta_{12}\to\delta_{c}\approx0.55$. At this critical value  we identify the roton wavevector $k_\text{rot}$ as the wavevector at which the roton first softens to zero energy  [see Fig.~\ref{excitationsfig0}(b)]. 
For $\delta_{12}>\delta_{c}$ the uniform state is unstable and the spin-stripe state is the ground state [cf.~Fig.~\ref{Fig:USPtransition}(b1), (b2), and (c1)].  The spin-stripe forms with $K=k_\mathrm{rot}$ at the critical point.

In Figs.~\ref{excitationsfig0}(a1) and (b1) we show the   dynamic structure factors for the $\delta_{12}=0.45$ case.  Excitations with $\lambda=+1$ and $\lambda=-1$ excitations contribute to the $S_+$ and $S_-$ dynamic structure factors, respectively, and the dip in $S_-$ clearly reveals the spin roton. 
In Fig.~\ref{Fig:PDiag}(a) the spin-roton softening is used to identify the roton stability boundary (light blue line).

 \subsubsection{Balanced spin-stripe state}
 
 \begin{figure}[htp!] 
 	\centering
 	\includegraphics[trim=0 0 0 0,clip=true,width=\linewidth]{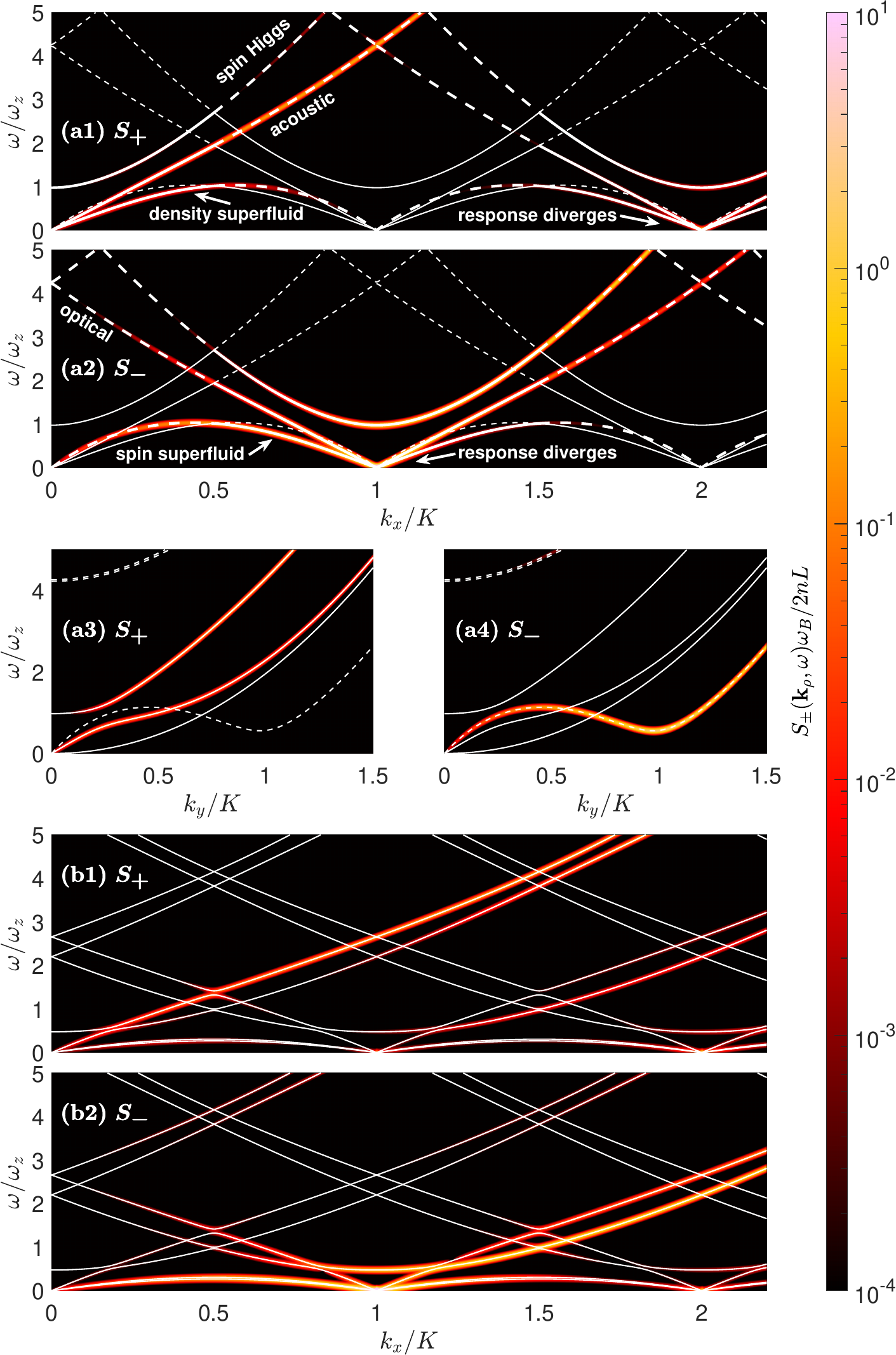}
 	\caption{
 		Excitations and dynamic structure factors for (a) balanced with $\mu_2^m/\mu_1^m = -1$, $\delta_{12}=0.6$
        and (b) unbalanced spin-stripe states with $\mu_2^m/\mu_1^m=0$, $\delta_{12}=1.15$. White lines indicate periodically extended excitation spectrum. Solid ($\lambda=1$) and dashed ($\lambda=-1$) lines in (a) indicate the associated symmetries for the balanced case. The dynamical structure factors were broadened as for Fig.~\ref{excitationsfig0}.
 	}\label{excitationdynamicalstructure}
 \end{figure}

In Figs.~\ref{excitationdynamicalstructure}(a1) and (a2) we consider the excitation spectrum (white lines) for $\delta_{12}>\delta_c$ where the ground state is a spin-stripe state [case shown in Fig.~\ref{Fig:USPtransition}(c1)]. Here we show the excitations propagating with quasi-momentum normal to the stripe (i.e.~along $x$). The excitations are repeated in the extended zone scheme to allow comparison with the dynamic structure factors.
We observe three gapless excitation branches in this phase, with two being $\lambda=1$ excitation branches, and one being  a $\lambda=-1$  branch. 
This is consistent with the prediction of $2+D$ Nambu-Goldstone modes for a two-component  supersolid state \cite{Watanabe2012a}, where $D$ is the dimensionality of crystalline order (here $D=1$ for the spin-stripe state). We denote the lowest energy $\lambda=1$ branch as the density superfluid mode (adopting the terminology used in Ref.~\cite{Kirkby2023a}) and the two upper branches as the (gapless) acoustic and (gapped) spin-Higgs modes, respectively. Similarly, we denote the lowest energy $\lambda=-1$ gapless branch as the spin superfluid mode and the upper gapped branch as the optical mode. For comparison there are two gapless modes and a Higgs mode in scalar dipolar supersolids \cite{Roccuzzo2019a}.

In Figs.~\ref{excitationdynamicalstructure}(a1) and (a2) the density and spin density dynamic structure factors (shaded colors) are shown. Here we observe that the excitation contributions alternate between causing density and spin fluctuations as $k_x$ increases and we cross Brillouin zones. For example, only the $\lambda=1$ modes (solid) contribute to $S_+$ for $0\le k_x\le K/2$, whereas only the $\lambda=-1$ modes (dashed) contribute for $K/2\le k_x\le 3K/2$. This arises from the  nonsymmorphic symmetry identified in Sec.~\ref{Sec:BalanceExcitations}, and can also be seen as consequences of Fourier transforming $L/2$-periodic and $L/2$-antiperiodic functions from Eqs.~\eqref{Eq:DenFlu_FFT} and \eqref{Eq:uv_nonsymm}. This symmetry also causes a degeneracy in the excitations, notably that pairs of $\lambda=1$ and $\lambda=-1$ bands are degenerate at the band edge\footnote{We have the additional symmetry operator $P_{k_x}$ describing the invariance of the system under an inverse of the $x$ component of momentum. This pins the band crossing to the edge of the Brillouin zone \cite{Zhao2016a}.} $q_x=K/2$. This degeneracy is most readily seen in the dynamic structure factor results where the $\lambda=\pm1$ excitation bands are plotted together for reference, as there is no sign of avoided crossings at the Brillouin zone boundaries. Another feature of this symmetry is that spin density response diverges at $k_x\to K$ and $\omega\to 0$ [see Fig.~\ref{excitationdynamicalstructure}(a2)] and the density response diverges at $k_x\to 2K$ and $\omega\to 0$ [see Fig.~\ref{excitationdynamicalstructure}(a1)].
 
In Figs.~\ref{excitationdynamicalstructure}(a3) and (a4) we examine the excitations propagating parallel to the stripe (i.e.~along $y$). In this direction momentum is a good quantum number. We see the three gapless bands have different behavior compared to the results for propagation normal to the stripe [cf.~Figs.~\ref{excitationdynamicalstructure}(a1) and (a2)]. 
Most notably, the lowest energy density band has quadratic (i.e.~free particle) character and corresponds to a transverse excitation of the stripe, and as a result makes no contribution to the density structure factor. Similarly, the higher density and spin bands that contribute to the respective structure factors, all correspond to longitudinal excitations of the stripes.

\subsubsection{Unbalanced case: dipolar-nondipolar mixture}
In Figs.~\ref{excitationdynamicalstructure} (b1) and (b2) we consider the excitations and structure factor for an unbalanced case of a dipolar-non-dipolar mixture  ($\mu_2^m=0$) for the stripe state shown in Fig.~\ref{Fig:USPtransition}(c2). Here we can no longer separate excitations into pure spin or density character (i.e.~$\lambda$ is no longer a good quantum number). The nonsymmorphic symmetry is broken for this case and the exact degeneracies at the band edge are now replaced by avoided crossings.

\section{Quench dynamics}\label{Sec:Qdynamics}

\begin{figure}[htp!] 
	\centering 
	\includegraphics[width=\linewidth]{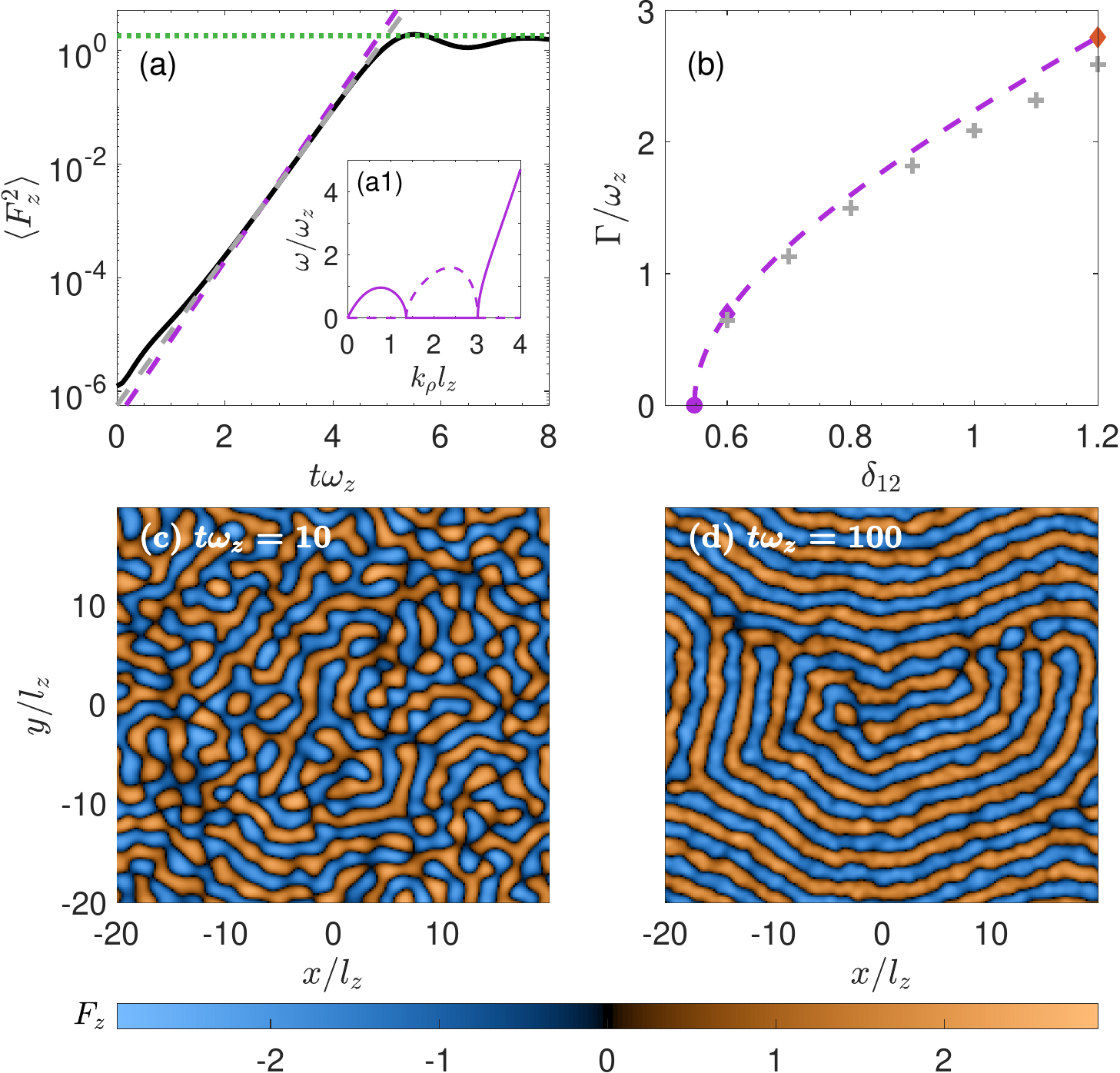}
    \caption{Early- and moderate-time growth of  spin-stripe order following a sudden quench to  $\delta_{12}=0.8$. (a) Evolution of $\langle F_z^2\rangle$ (black line).  An exponential growth rate from fitting to  data (gray-dashed line) and  Bogoliubov prediction  $\sim e^{4\Gamma t}$ (purple-dashed). Inset (a1) shows real (solid) and imaginary (dashed) $\lambda=-1$ excitation spectrum. (b) Comparison of  Bogoliubov prediction (purple-dashed) to the fitted exponential growth rate (gray crosses) for quenches to various $\delta_{12}$ values. The three colored markers in (b) have the parameters indicated by the same markers in Fig.~\ref{Fig:PDiag}(a). Examples of the (c) early   ($t=10/\omega_z$) and (d) moderate  ($t=100/\omega_z$) time spin density patterns for $\delta_{12}=0.8$. Other parameters are the same as the values used in Fig.~\ref{Fig:USPtransition}(c1). The green line in (a) indicates   $\langle F_z^2\rangle$  for the spin-stripe ground state. Simulations performed on a square grid with a side length $l=40l_z$, $200$ points in each direction, and periodic boundary conditions.  \label{earlytime}}
\end{figure}

We now examine the dynamics of a quench from the uniform miscible to spin stripe state to explore the dynamics of stripe formation. Specifically, at time $t=0$ we set $\delta_{12}$ to a value where a spin-stripe state is expected and simulate the dynamics according to the time-dependent GPE. This implements an instant quench in the interspecies interaction.

We  perform these simulations on a square grid of side length $l$ with periodic boundary conditions.  We add a complex normally distributed noise on to the initial miscible state $\psi_i=1$, which mimics quantum and thermal fluctuations in the quantum field theory \cite{Saito2007a,cfieldRev2008,Barnett2011a}. The noise  is added to momentum space to modes with wavevectors  $k_\rho l_z\leq 3.5$. This choice restricts the noise to the low kinetic energy modes, but ensures that the dynamically unstable Bogoliubov modes have some finite initial occupations. The noise is weak and changes the relative normalization of the  initial field  by $\sim 10^{-6}$. 
 
Immediately following the quench unstable modes initiate the immiscibility dynamics. The  immiscibility is conveniently described using the (pseudo) spin density
\begin{align}
	F_z(\brho)=|\psi_1(\brho)|^2-|\psi_2(\brho)|^2,
\end{align}
which characterizes the differences in densities of both components. To quantify the early-time dynamics of the spin-stripe formation we use the expectation
\begin{align}
	\langle F_z^2\rangle &\equiv \frac1A\int  d\brho\,F_z^2(\brho), 
\end{align}
where $A=l^2$ is the total area of the system.
  The evolution of $ \langle F_z^2\rangle$ following the quench is shown in Fig.~\ref{earlytime}(a), where $ \langle F_z^2\rangle$ has a small non-zero value at $t=0$ arising from the initial noise, and grows exponentially with time, which reflects the unstable excitations of the initial state. We then identify the growth rate of $\langle F_z^2\rangle$ associated with the most unstable mode as
 \begin{align}
     \Gamma = \max_\bnu\mathrm{Im}\left\{\omega_\bnu\right\}.\label{GammaGrowth}
 \end{align}
The results in Fig.~\ref{earlytime}(a)  show that this provides a good quantitative description of the exponential growth at early times. In  Fig.~\ref{earlytime}(b) we consider how the early time growth rate changes with $\delta_{12}$ and verify the applicability of Eq.~(\ref{GammaGrowth}) for these cases. This growth eventually saturates  on  a time scale of $t\sim 10/\omega_z$, when $\langle F_z^2\rangle$ saturates to a value close to the expected value for the spin-stripe ground state.   

 In Fig.~\ref{earlytime}(c) we show an early-time spin density pattern  following the quench. Later, at $t=100/\omega_z$ [see Fig.~\ref{earlytime}(d)], the spin-stripe structure is already well-established, although the orientation of the stripes varies over space. From these stripes we can extract the mean reciprocal lattice vector, which we display as the black cross markers in Fig.~\ref{Fig:USPtransition}(b1), and are seen to be in good agreement with the expected ground state reciprocal lattice vectors.
 
 The early-time growth in immiscible binary and spin-1 condensates has been considered  in previous work (e.g.~see Refs.~\cite{Hofmann2014a,Williamson2016a,Williamson2016b}) and the initial evolution of $\langle F_z^2\rangle$ is similar to our observations. However, in these non-dipolar systems irregular domains develop, rather than regular spin-stripes.  As time progresses  the spin domains grow in size with time in a coarsening process \cite{Bray1994a} that evolves the system towards one large domain of each component (also see \cite{Guzman2011a,Kudo2013a,Takeuchi2015a,Bourges2017a,Takeuchi2018a}).

\begin{figure}[htp!] 
	\centering
	\includegraphics[width=\linewidth]{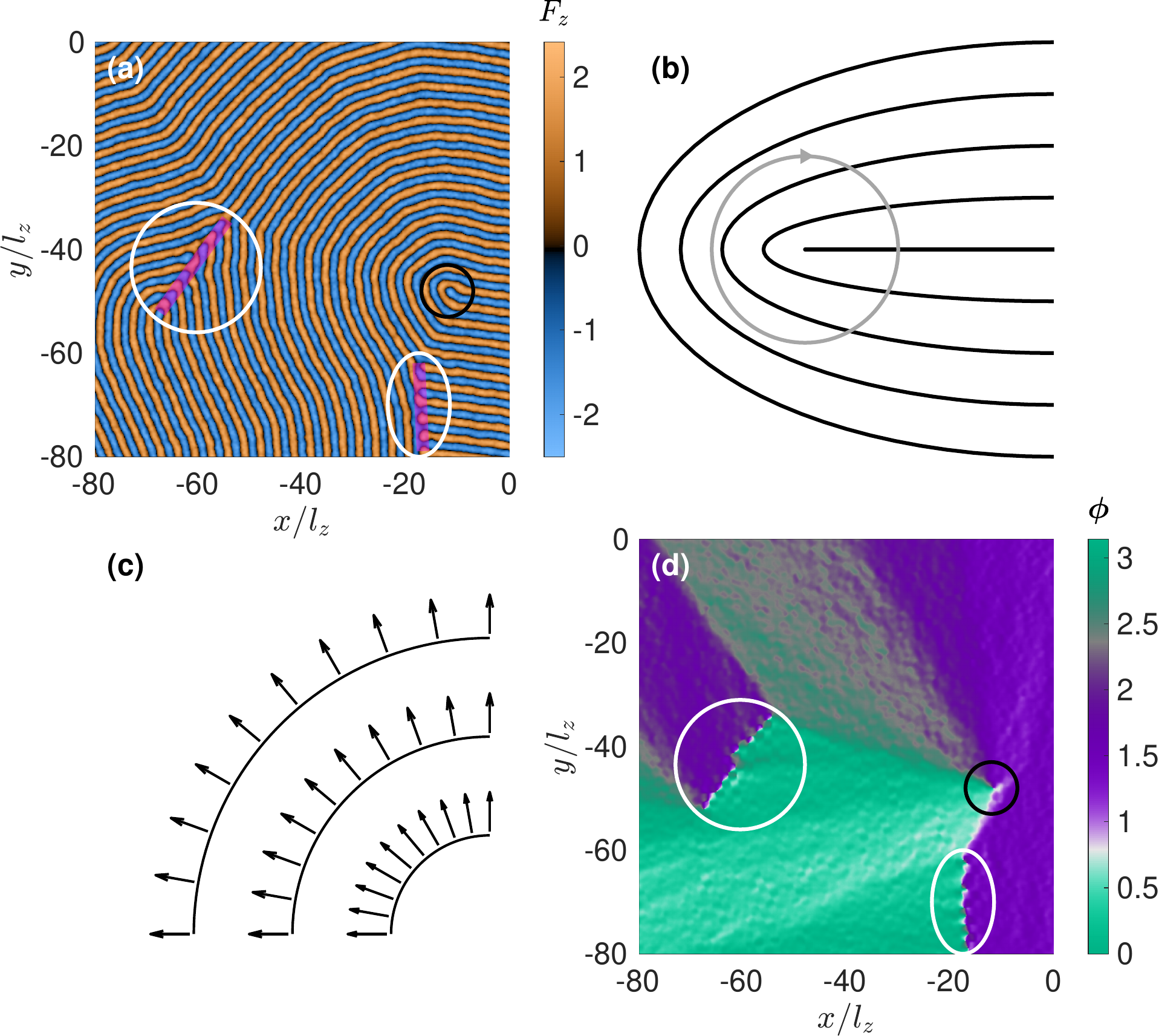}
    \caption{(a) Illustrative example of the spin-density at $t=1600/\omega_z$ following a quench to the spin-stripe phase. The black circle identifies a disclination defect in the stripe pattern and the two white circles identify grain boundary defects.
	(b) Schematic of a $+1/2$ disclination defect in a stripe pattern. The orientation of a vector perpendicular to the stripes (black lines) [see (c)] rotates by 180$^\circ$   as we travel over the closed circular path drawn around the core of the defect.
(c) Schematic example of the normal vectors $\hat{\mathbf{n}}$ perpendicular to the stripes (also see text). We use these normal vectors to quantify the orientation order. (d) The orientation order (angle) of the normal vectors corresponds to the spin density in (a). Other parameters are the same as the values used in Fig.~\ref{earlytime}(a). Simulation is performed on a square grid with a side length $l = 160l_z$ (one quarter of the simulation area is shown), with $800$ points in each direction.
	\label{ssOrderDefects}}
\end{figure}

\begin{figure}[htp!] 
	\centering
	\includegraphics[width=\linewidth]{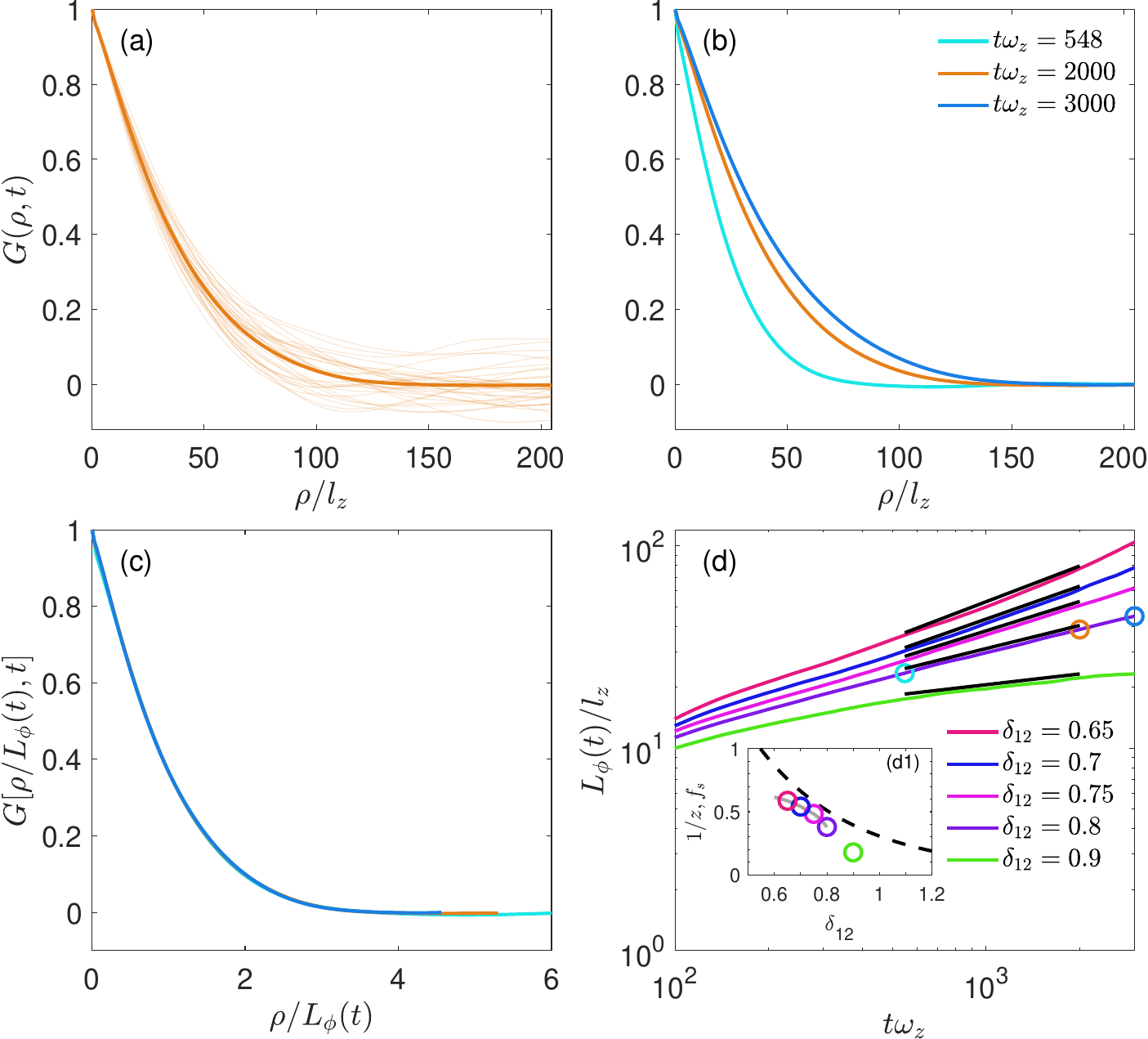}
    \caption{Phase ordering kinetics of the spin-stripe state. (a) Spatially and angular averaged correlation function for 32 trajectories (thin lines) and the averaged result (thick line) at $t=2000/\omega_z$ after quench. (b) The orientation order parameter correlation function at three different times after the quench.
(c) Collapse of correlation functions in (b) by scaling with the correlation length $L_\phi(t)$.
(d)  Growth of the correlation length at late times following the quench indicating a power law growth. The circles mark the results corresponding to the cases shown in (b). We used $32$ trajectories for $\delta_{12}<0.9$, and $8$ trajectories for $\delta_{12}=0.9$. Inset: comparison of dynamic critical exponents  (circles) and the superfluid fraction $f_s$ (dashed line).
Other parameters are the same as the values used in Fig.~\ref{earlytime}. Simulations performed on a square grid with a side length $l=409.6l_z$, 2048 points in each direction. 
	\label{ssOrdering}}
\end{figure}

We are also interested in how the orientation order grows in the spin stripe phase. Initially the stripe domains (i.e.~regions with regular stripe spacing and orientation) extend over small length scales [e.g.~see Fig.~\ref{earlytime}(c)]. At later times the   domains are observed to be  larger, e.g.~compare Figs.~\ref{earlytime}(d) and \ref{ssOrderDefects}(a).  We can see two types of defect in the example presented in Fig.~\ref{ssOrderDefects}. First, we can see a disclination point defect indicated by the black circle drawn in Fig.~\ref{ssOrderDefects}(a) [also see Fig.~\ref{ssOrderDefects}(b)]. Second, we observe a grain boundary line defect indicated by the thick magenta lines inside the white circles in Fig.~\ref{ssOrderDefects}(a). These defects are known from the theory of 2D solids, e.g.~see  Ref.~\cite{Nelson1979a}.

To characterize the spin-stripe orientation order, the relevant order parameter is the orientation angle $\phi(\brho)$ of the normal vectors of the spin-density in the $xy$-plane \cite{Harrison2002a}.  We show a schematic example of the unit normal vectors $\hat{\mathbf{n}}$  in Fig.~\ref{ssOrderDefects}(c)\footnote{The stripe orientation order can be  characterized by a nematic director. To reflect this we take the normal vectors to be in the upper half-plane $\hat{\mathbf{n}}=\mathrm{sign}\left(\frac{\partial F_z}{\partial y}\right)\frac{\bm{\nabla}F_z}{|\bm{\nabla}F_z|}$.} to demonstrate how we map the spin density in Fig.~\ref{ssOrderDefects}(a) onto $\phi(\brho)$ in Fig.~\ref{ssOrderDefects}(d). Because the magnitude of the normal vectors vanish at the local maximum and minimum of $F_z$, and due to fluctuations, we remove  some short wavelength noise by a Gaussian filtering. Some residual noise remains on the filtered order parameter but generally it is seen to vary smoothly over space. The results  of Fig.~\ref{ssOrderDefects}(d) also clearly reveal the point and line defects.

 To examine how order develops after the quench we evaluate the order parameter correlation function
 \begin{align}
     G(\brho,t)=\frac1A\int d\brho' \,e^{2i[\phi(\brho+\brho')-\phi(\brho')]}.
 \end{align}
 In practice the result is calculated using angular averaging to produce $ G(\rho,t)$. An example of the late-time correlation function for 32 independent quench trajectories are shown in  Fig.~\ref{ssOrdering}(a) as thin lines, and the average over these trajectories as a thick line.  Using this technique, we construct the averaged correlation functions at different times [see Fig.~\ref{ssOrdering}(b)]. At each time we identify the correlation length $L_\phi(t)$ (i.e.~typical domain size) as where $G[L_\phi(t),t] = G(0,t)/e$. 
  The results in Fig.~\ref{ssOrdering}(c) show that the correlation functions scaled by their correlation lengths, i.e.~$G\left[\rho/L_\phi(t),t\right]$ exhibit a collapse to a time independent function.  This verifies dynamic scaling of the order parameter at late times in the phase ordering.
Here, the results shown are taken at times from $t\sim10^2/\omega_z$ to $\sim3\times10^3/\omega_z$. The lower time limit is to ensure that the correlation length is sufficiently large compared to the microscopic lengths (e.g.~stripe wavelength).  The upper time limit is imposed to prevent the correlation length becoming comparable to the grid extent, i.e.~we restrict the data to cases where $L_\phi(t)\lesssim l/4$. 
 
In Fig.~\ref{ssOrdering}(d) we show several averaged correlation length $L_\phi(t)$ growth curves for quenches to  $\delta_{12}$ values in the range $0.65\leq\delta_{12}\leq0.9$.  For $0.65\leq\delta_{12}\leq0.8$ these are seen to grow with time consistent with a power law $L_\phi\sim t^{1/z}$, with an exponent $1/z$ that depends on the values of $\delta_{12}$ [see fits in Fig.~\ref{ssOrdering}(d), and exponents shown in the inset].  The result for $\delta_{12}=0.9$ grows more slowly and the domain size appears to saturate at late times. Our results for $\delta_{12}=1$ and $1.2$ (not shown) are similar to $\delta_{12}=0.9$ case, but once the domain size reaches $L_\phi\sim10l_z$, any further increase is very slow.
  
Several factors may play a role in the phase ordering dependence on $\delta_{12}$. First,  as $\delta_{12}$ increases, the density contrast of each component increases and transport is inhibited. Indeed, the broken translational invariance of the spin-stripe state leads to a reduction in the superfluid fraction of each component, which we quantify using the Leggett upper bound \cite{Leggett1970a} on component 1 [see inset to Fig.~\ref{ssOrdering}(d)]. We observe that the dynamic critical exponent decreases as the superfluid fraction decreases.  A second factor is that for deeper quenches (i.e.~to large $\delta_{12}$) a higher density of disclinations is created. Studies of stripe pattern ordering in 2D smectic systems found that as the disclination density increased, the phase ordering progressed at a significantly slower rate
 \cite{Harrison2002a}.

\section{Conclusions}\label{Sec:Conc}

In this work we have characterized the phase diagram for a quasi-2D  binary  BEC focusing on the spin-stripe state, which occurs when there is a difference in the DDIs between the components.  We have also calculated the excitations of the spin-stripe state, and explored the role of $\mathbb{Z}_2$ and nonsymmorphic symmetries in the balanced regime. This work laid the foundation for us to study the dynamics of how stripe order forms following a sudden quench from an immiscible state into the parameter regime where the spin-stripe state is the expected ground state. The system exhibits novel dynamics in this transition as domains consisting of spin-stripes with different orientations form across the system. At the interface between these domains, various defects such as grain boundaries and disclinations occur. Our results showed that as time evolves,  dynamic scaling  can occur, although for larger $\delta_{12}$ values we find that the domains are almost frozen and grow very slowly.
For values  of $\delta_{12}$ where phase ordering occurs, we found dynamic critical exponents in the range $1/z\sim 0.4$ -- $0.6$. 
For comparison, work on binary fluids \cite{Furukawa1985a} has established grow laws of $t^{1/3}$, $t$ and $t^{2/3}$ in the diffusive, viscous hydrodynamic and inertial hydrodynamic regimes, respectively \cite{Furukawa1985a,Bray1994a}. 
The 2D smectic liquid crystal also has a stripe pattern, although it is a single component system. Experimental studies of the phase ordering in 2D smectics   \cite{Harrison2000a,Harrison2002a}  found a $t^{1/4}$ growth law, reduced from the $t^{1/2}$ law observed  in  2D  nematic liquid crystals.  In the smectic system the growth of order was observed to follow the average spacing between disclinations. Furthermore, we find that the phase ordering changes character as $\delta_{12}$ increases and the number of defects increases. For sufficiently high $\delta_{12}$ values the phase ordering appears to stop and microscopic domains remain frozen in.
This suggests that a detailed investigation of the defect dynamics could shed light on the ordering dynamics in the spin-stripe phase.

Following on the experimental progress in producing binary dipolar mixtures, our predictions, particularly for the ground state structure and excitations, could be explored with these systems using quasi-2D box potentials \cite{Navon2021a}.    
Experimental work on the spin-stripe phase transition dynamics would complement efforts looking at the ordering dynamics of immiscible quantum phases explored in spinor BECs.  Of note is the experiment of Huh \textit{et al.}~\cite{Huh2023a} that observed the ordering dynamics in the easy-axis regime of a ferromagnetic spin-1 BEC, where the system evolves as an immiscible two-component mixture.  Those experiments were able to find universal scaling in the evolution of the spin correlations (cf.~Ref.~\cite{Prufer2018a}).   However, even in smaller systems the early stage evolution of the transition to a spin-stripe could be studied. In addition to the grow rates of local order, it would be interesting quantify the emergence  and evolution of defects.

\noindent {\bf Acknowledgments: $\,$}
We would like to thank Philip Brydon for useful discussions. We acknowledge support from the Marsden Fund of the Royal Society of New Zealand and the contribution of NZ eScience Infrastructure (NeSI) high-performance computing facilities.


%

\end{document}